%% file: ppDijetFwdSumETPaper.tex
\newcommand*{\ATLASLATEXPATH}{}

\documentclass[cernpreprint,texlive=2011,txfonts,UKenglish]{atlasdoc} 

\pdfoutput=1

\usepackage [compress,comma,numbers]{natbib}
\usepackage[biblatex=false]{atlaspackage}

\makeatletter
\def\NAT@def@citea{\def\@citea{\NAT@separator}}
\makeatother

\usepackage{\ATLASLATEXPATH atlasphysics}

\usepackage{ppDijetFwdSumETPaper-defs}

\newcommand{\paper}{Letter}

\hypersetup{pdftitle={ATLAS draft},pdfauthor={The ATLAS Collaboration}}

\input{ppDijetFwdSumETPaper-metadata}

\begin{document}

\maketitle

\tableofcontents

\section{Introduction}

Properties of the underlying event at large rapidity in proton--proton
(\pp) collisions in the presence of a hard parton--parton scattering
are sensitive to many features of hadronic interactions. Previous
studies of the underlying event mainly focused on probing the region
transverse to final-state jets at
mid-rapidity~\cite{PhysRevD.65.092002,Aaltonen:2010rm,Aad:2014hia,Aad:2012mfa}. This
\paper\ presents a study of the transverse energy produced at small
angles with respect to the proton beam, a region where particle
production may be particularly sensitive to the colour connections
between the hard partons and the beam remnants. Such measurements are
needed to constrain particle production models, which systematically
underpredict the total transverse energy at forward rapidities in
hard-scattering events~\cite{Aad:2012mfa}.

Measurements of transverse energy production at large rapidity are
also needed to aid in the interpretation of recent results on jet
production in proton--lead ($p$+Pb)
collisions~\cite{ATLAS:2014cpa,Chatrchyan:2014hqa}. In these
collisions, hard scattering rates are expected to grow with the
increasing degree of geometric overlap between the proton and the
nucleus. Simultaneously, the level of overlap is traditionally thought
to be reflected in the rate of soft particle production, particularly
at large pseudorapidity in the nucleus-going direction. The recent
results found that single and dijet production rates in the
proton-going (forward, or projectile) direction are related to the
underlying-event activity in the nucleus-going (backward, or target)
direction in a way that contradicts the models of how jet and
underlying-event production should correlate. Specifically, the
average transverse energy produced in the backward direction was found
to systematically decrease, relative to that for low-energy jet
events, with increasing jet energy. This decrease resulted in an
apparent enhancement of the jet rate in low-activity, or peripheral,
events and a suppression of the jet rate in high-activity, or central,
events.

These results have several competing interpretations. For example,
they are taken as evidence that proton configurations with a parton
carrying a large fraction $x$ of the proton longitudinal momentum
interact with nucleons in the nucleus with a significantly smaller
than average cross-section~\cite{Alvioli:2014eda}. Alternatively,
other authors have argued that in the constituent nucleon--nucleon
($NN$) collisions, energy production at backward rapidities naturally
decreases with increasing $x$ in the forward-going proton, either
through the suppression of soft gluons available for particle
production~\cite{Bzdak:2014rca} or from a rapidity-separated
energy-momentum conservation between the hard process and soft
production~\cite{Armesto:2015kwa}. More generally, the modification of
soft particle production in $NN$ collisions in the presence of a hard
process is expected to affect estimates of the collision geometry of
\pPb collisions with a hard
scatter~\cite{Adare:2013nff,Perepelitsa:2014yta,Adam:2014qja}. Thus a
control measurement in \pp\ collisions to determine how soft particle
production at negative pseudorapidities varies with the $x$ in the
projectile (target) beam-proton headed towards positive (negative)
rapidity can provide insight into the relevance of these various
scenarios.

This \paper\ presents a measurement of the average of the sum of the
transverse energy at large pseudorapidity\footnote{ATLAS uses a
  right-handed coordinate system with its origin at the nominal
  interaction point (IP) in the centre of the detector and the
  $z$-axis along the beam pipe. The $x$-axis points from the IP to the
  centre of the LHC ring, and the $y$-axis points upward. Cylindrical
  coordinates $(r,\phi)$ are used in the transverse plane, $\phi$
  being the azimuthal angle around the beam pipe. The pseudorapidity
  is defined in terms of the polar angle $\theta$ as
  $\eta=-\ln\tan(\theta/2)$.}, \meansumET, downstream of one of the
protons in $pp$ collisions, as a function of the hard-scattering
kinematics in dijet events. For each kinematic selection, \meansumET
is the average of the \sumET distribution in the selected events. The
\sumET measurement was deliberately made in only one of the two
forward calorimeter modules on either side of the interaction
point. This was done in analogy with the centrality definition in
$p$+Pb collisions~\cite{ATLAS:2014cpa,Aad:2015zza}, which is
characterised by the \sumET in the forward calorimeter module situated
at $-4.9 < \eta < -3.2$, in the nucleus-going direction. In
\pp\ collisions the asymmetric choice of the \sumET-measuring region
means that the target proton plays the role of one of the nucleons in
the Pb nucleus.

The value of \sumET was measured by summing the transverse energy in
the forward calorimeter cells and correcting for the detector
response. The average value, \meansumET, is reported as a function of
the average dijet transverse momentum, $\pTavg = (p_\mathrm{T,1} +
p_\mathrm{T,2})/2$, and pseudorapidity, $\etadijet = (\eta_{1} +
\eta_{2})/2$. In these quantities, $p_\mathrm{T,1}$ and $\eta_{1}$ are
the transverse momentum and pseudorapidity of the leading
(highest-\pT) jet in the event, while $p_\mathrm{T,2}$ and $\eta_{2}$
are those for the subleading (second highest-\pT) jet. Results are
also reported as a function of two kinematic quantities \tildexone and
\tildextwo defined by

\begin{eqnarray}
\tildexone & = & \pTavg ( e^{+\eta_1} + e^{+\eta_2})/\sqrt{s}, \\
\tildextwo & = & \pTavg ( e^{-\eta_1} + e^{-\eta_2})/\sqrt{s}.
\end{eqnarray}

In a perturbative approach, at leading order, $\tildexone$
($\tildextwo$) corresponds approximately to the Bjorken-$x$ of the
hard-scattered parton in the beam-proton with positive (negative)
rapidity. Estimates of the initial parton--parton kinematics through
jet-level variables have been used previously in dijet measurements at
the CERN S$p\bar{p}$S collider~\cite{Arnison:1983dk,Bagnaia:1984ip}
and in measurements of dihadrons in $d$+Au collisions at
RHIC~\cite{Adare:2011sc}. Finally, to better reveal the relative
dependence of \meansumET on the hard-scattering kinematics, results
are also reported as a ratio to a reference value \meansumETref, which
is the \meansumET evaluated at a fixed choice of dijet kinematics,
$50$~\GeV $< \pTavg < 63$~\GeV\ and $\left|\smash{\etadijet}\right| <
0.3$.

\begin{figure}[!p]
\centering
\includegraphics[width=0.65\columnwidth]{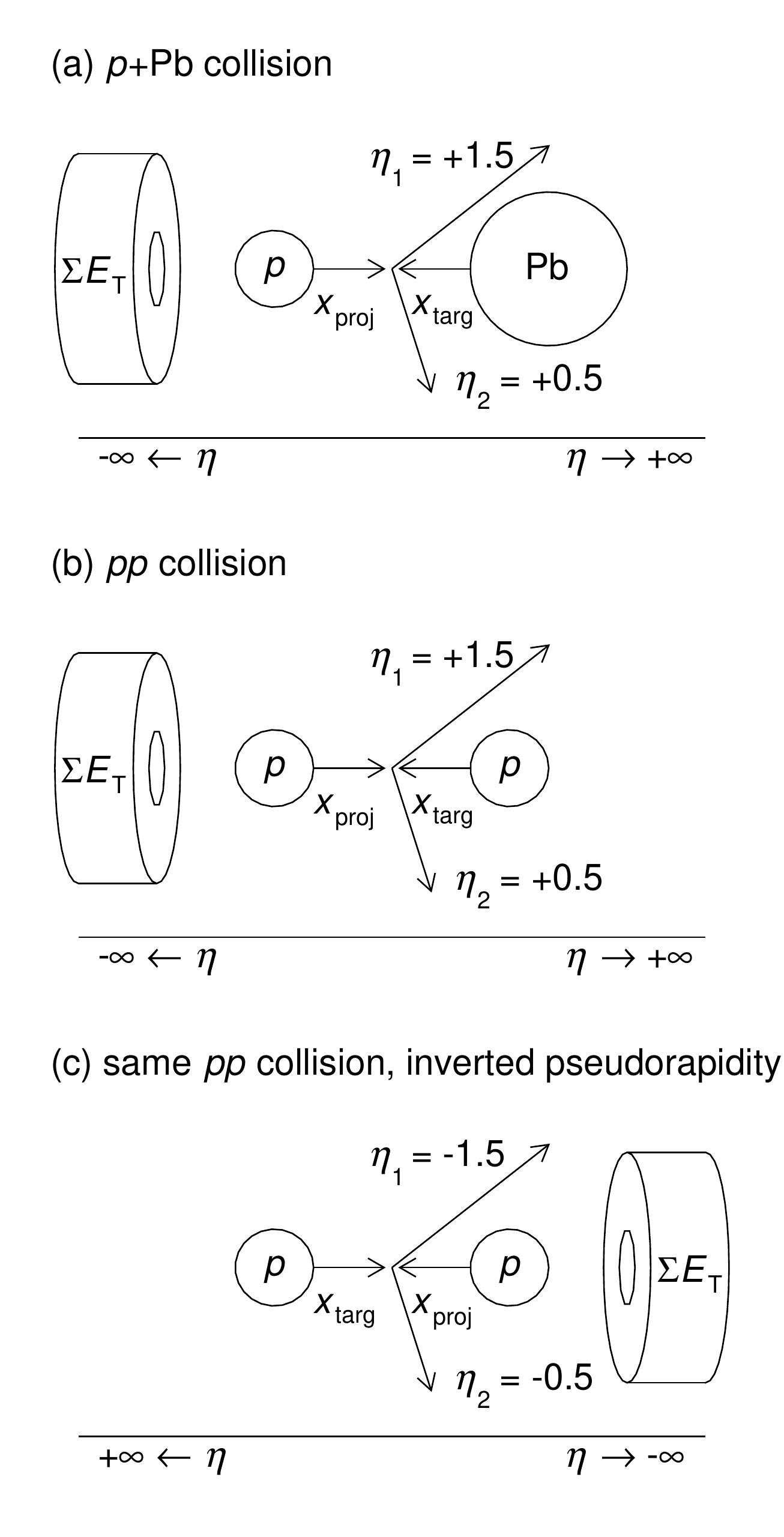}
\caption{\label{fig:Schematic} Schematic illustration of the kinematic
  variables in the measurement. Panel (a) illustrates the convention
  in \pPb collisions. Panels (b) and (c) illustrate how a single $pp$
  event provides a measurement of \sumET at two values of $\etadijet =
  (\eta_{1} + \eta_{2})/2$, in this case for $\etadijet = +1$ and for
  $\etadijet = -1$, respectively. }
\end{figure}

Fig.~\ref{fig:Schematic} schematically illustrates the meaning of the
kinematic variables utilised in this measurement. The top panel in
Fig.~\ref{fig:Schematic} shows the convention used in $p$+Pb
collisions at ATLAS, in which the proton beam is the ``projectile''
and has positive rapidity, while the nuclear beam is the ``target''
and has negative rapidity. The centrality of the $p$+Pb collision, an
experimental quantity sensitive to the collision geometry, is
characterised by the \sumET in the forward calorimeter situated in the
nucleus-going direction. The middle panel in Fig.~\ref{fig:Schematic}
illustrates the measurement in $pp$ collisions reported in this
\paper, in which the proton beam with positive rapidity is considered
to be the analogue of the projectile proton in $p$+Pb collisions,
while the target proton with negative rapidity is the analogue of a
single nucleon within the Pb nucleus, and the \sumET is measured in
the forward calorimeter downstream of the target proton. Due to the
symmetric nature of $pp$ collisions, each event can also be
interpreted by exchanging the roles of the target and projectile
between the two protons, and measuring the \sumET in the opposite
forward calorimeter module. To keep the same convention in this case,
the $z$-axis (and thus the pseudorapidity) is inverted and the
kinematic variables are determined within this new coordinate system
as shown in the bottom panel of Fig.~\ref{fig:Schematic}. The full
analysis was performed separately using each forward calorimeter side,
one at a time, and the final results were obtained by averaging the
\meansumET measurements from each side. This increased the number of
\sumET measurements by a factor of two and also provided an important
cross-check on the detector energy scale. For simplicity, all $\eta$
values in the selection cuts and \etadijet values in the results
described below are always presented according to the convention where
\sumET is measured at negative pseudorapidity.

The dataset used in this measurement was collected during the
$\sqrt{s}=2.76$~\TeV\ \pp\ collision data-taking in February 2013 at
the Large Hadron Collider, with an integrated luminosity corresponding
to $4.0$~pb$^{-1}$. During data-taking, the mean number of $pp$
interactions per bunch crossing varied from $0.1$ to $0.5$. This
dataset is particularly suitable for the measurement because the small
mean interaction rate per crossing allows rejection of dijet-producing
$pp$ events with additional $pp$ interactions in the same bunch
crossing (pileup) with good systematic control while simultaneously
having enough integrated luminosity to measure dijet production over a
wide kinematic range with good statistical precision.

\section{Experimental setup}

The ATLAS detector is described in detail in
Ref.~\cite{Aad:2008zzm}. This analysis uses primarily the tracking
detectors, the calorimeter, and the trigger system. Charged-particle
tracks were measured over the range $\left|\eta\right| < 2.5$ using
the inner detector, which is composed of silicon pixel detectors in
the innermost layers, silicon microstrip detectors, and a straw-tube
transition-radiation tracker ($\left|\eta\right| < 2.0$) in the outer
layer, all immersed in a $2$~T axial magnetic field. The calorimeter
system consists of a liquid argon (LAr) electromagnetic calorimeter
($\left|\eta\right| < 3.2$), a steel/scintillator sampling hadronic
calorimeter ($\left|\eta\right| < 1.7$), a LAr hadronic calorimeter
($1.5 < \left|\eta\right| < 3.2$), and a forward calorimeter ($3.2 <
\left|\eta\right| < 4.9$). The forward calorimeter is composed of two
modules situated at opposite sides of the interaction region and
provides the \sumET measurement. The modules consist of tungsten and
copper absorbers with LAr as the active medium, which together provide
ten interaction lengths of material, and are segmented into one
electromagnetic and two hadronic sections longitudinal in the shower
direction. The 1782 cells in each forward calorimeter module are
aligned parallel to the beam axis and therefore are not projective,
but have a segmentation corresponding to approximately $0.2\times0.2$
in $\eta$ and $\phi$.

Data were acquired for this analysis using a series of central-jet
triggers covering $\left|\eta\right| < 3.2$ with different
(increasing) jet-\pT thresholds, ranging from $40$~\GeV\ to
$75$~\GeV~\cite{Aad:2012xs}. Each trigger was prescaled, meaning that
only a fraction of events passing the trigger criteria were
ultimately selected, and these fractions varied with time to
accommodate the evolution of the luminosity within an LHC fill. This
fraction increased for triggers with increasing jet \pT threshold and
the highest-threshold trigger, which dominates the kinematic range
studied in this \paper, sampled the full integrated luminosity.

\section{Monte Carlo simulation}

Monte Carlo (MC) simulations of $\sqrt{s} = 2.76$~\TeV\ $pp$
hard-scattering events were used to understand the performance of the
ATLAS detector, to correct the measured $\sumET$ and dijet kinematic
variables for detector effects, and to determine the systematic
uncertainties in the measurement. Three MC programs were used to
generate event samples with the leading-jet \pT in the range from
$20$~\GeV\ to $1$~\TeV: the \PYTHIA 6
generator~\cite{Sjostrand:2006za} with parameter values chosen to
reproduce data according to the {\tt AUET2B} set of tuned parameters
(tune)~\cite{ATL-PHYS-PUB-2011-009} and {\tt CTEQ6L1} parton
distribution function (PDF) set~\cite{Pumplin:2002vw}; the \PYTHIA 8
generator~\cite{Sjostrand:2007gs} with the {\tt AU2}
tune~\cite{ATL-PHYS-PUB-2012-003} and {\tt CT10} PDF
set~\cite{Guzzi:2011sv}; and the \HERWIG generator~\cite{Bahr:2008pv}
with the {\tt UE-EE-3} tune~\cite{Gieseke:2012ft} and {\tt CTEQ6L1}
PDF set. The generated events were passed through a full \GEANT4
simulation~\cite{Agostinelli:2002hh,Aad:2010ah} of the ATLAS detector
under the same conditions present during data-taking. The simulated
events included contributions from pileup similar to that in data.

At the particle level, jets are defined by applying the anti-$k_{t}$
algorithm~\cite{Cacciari:2011ma} with radius parameter $R$ of $0.4$ to
primary particles\footnote{Primary particles are defined as
  final-state particles with a proper lifetime greater than $30$~ps.}
within $\left|\eta\right| < 4.9$, excluding muons and
neutrinos. \sumET is defined at the particle level as the sum of the
transverse energy of all primary particles within $-4.9 < \eta <
-3.2$, including muons and neutrinos, and with no additional kinematic
selection.

\section{Event reconstruction and calibration}

The vertex reconstruction, jet reconstruction and calibration, and
\sumET measurement and calibration procedures are described in this
section. They were applied identically to the experimental data and
the simulated events.

\subsection{Track and vertex reconstruction}

In the offline analysis, charged-particle tracks were reconstructed in
the inner detector with an algorithm used in previous measurements of
charged-particle multiplicities in minimum-bias
\pp\ interactions~\cite{Aad:2010ac}. Analysed events were required to
contain a reconstructed vertex, formed by at least two tracks with
$\pT > 0.1$~\GeV~\cite{ATLAS-CONF-2010-069}. The contribution from
pileup interactions was suppressed by rejecting events containing more
than one reconstructed vertex with five or more associated
charged-particle tracks. This requirement rejected approximately $8$\%
of events.

\subsection{Jet reconstruction and calibration}

The jet reconstruction and associated background determination
procedures closely follow those developed within ATLAS for jet
measurements in heavy-ion and $pp$
collisions~\cite{ATLAS:2014cpa,Aad:2012vca,Aad:2014wha,Aad:2014bxa}.
This procedure is summarised in the following and is described in more
detail in Ref.~\cite{Aad:2012vca}. Jets were reconstructed by applying
the anti-$k_{t}$ algorithm with $R=0.4$ to calorimeter cells grouped
into towers of size $\Delta\eta\times\Delta\phi=0.1\times0.1$. The
procedure provided an $\eta$- and sampling layer-dependent estimate of
the small energy density deposited by the soft underlying event from
pileup interactions in each crossing. The energies of the cells in
each jet were corrected for this estimate of the soft pileup
contribution. The \pT of the resulting jets was corrected for the
calorimeter energy response through a simulation-derived calibration,
with an additional {\em in situ} correction, typically at the percent
level, derived through comparisons of boson--jet and dijet \pT balance
in collision data and simulation~\cite{ATLAS-CONF-2015-016}.

\subsection{Forward transverse energy measurement and calibration}

The \sumET quantity was evaluated by measuring the sum of the
transverse energy in the cells in one forward calorimeter module
(\sumETraw). The energy signals from the cells were included in the
sum without any energy threshold requirement. This quantity was
corrected event-by-event to account for the detector response, using a
calibration procedure derived in simulation, to give an estimate of
the full energy deposited in the calorimeter (\sumETcalib). \PYTHIA 8
was found to give the best overall description of \sumET production
and of its dependence on dijet kinematics in data. Thus, a subset of
\PYTHIA 8 events with good kinematic overlap with the data and a wide
range of \sumET values was used to calibrate \sumETraw. The
calibration was derived by requiring that for each subset of simulated
events with a narrow range of particle-level \sumET values
(\sumETgen), the mean value of the \sumETcalib distribution in those
events corresponded to the mean value of $\sumETgen$. First, to
determine the average offset in the response ($\Delta$), the average
\sumETraw as a function of \sumETgen was extrapolated with a linear
fit to zero \sumETgen. This additive offset, which described the
average net effect of energy inflow from outside and energy outflow
from inside the fiducial pseudorapidity acceptance of $-4.9 < \eta <
-3.2$, was found to be approximately $\Delta \approx -0.7$~\GeV. It
also reflected the residual contribution from pileup interactions and
the average distortion of the signal from energy deposited by
collisions in previous bunch crossings. Second, the average response
($C$) was determined by the ratio of the mean offset-corrected
\sumETraw to each corresponding value of \sumETgen, $C =
\left<\smash{\sumETraw-\Delta}\right>/\sumETgen$. $C$ was found to be
approximately $0.7$ and varied only weakly with \sumETgen after the
offset correction. This residual dependence was modelled by evaluating
$C$ in narrow bins of \sumETraw and fitting the results with a smooth
function to produce a continuous interpolation, $C(\sumETraw)$. The
calibrated quantity in each event was determined by correcting the raw
quantity for the offset and average response, $\sumETcalib =
(\sumETraw - \Delta)/C(\sumETraw)$.

The closure of this calibration, defined as the ratio
$\left<\smash{\sumETcalib}\right>/\sumETgen$ as a function of
\sumETgen, was within $1$\% of unity. The closure was also checked for
different selections on the dijet kinematics, which have a variety of
$\mathrm{d}N/\mathrm{d}\!\sumET$ and
$\mathrm{d}\!\sumET/\mathrm{d}\eta$ distributions, by comparing the
mean \sumETcalib to the mean of the \sumETgen distribution in these
events. For the selections on dijet kinematics in which the closure
was statistically meaningful, it was within a few percent of unity and
the non-closure was accounted for in the systematic uncertainty
described below.

\section{Event selection and data analysis}

In the offline analysis, the leading jet in each jet-triggered event
was required to match to a jet reconstructed at the trigger
stage. Only one trigger was used for each leading-jet \pT
interval. This trigger was chosen to be the one with the highest
integrated luminosity that was simultaneously $>99$\% efficient within
the interval. The contribution from each event to the \sumET
measurement was weighted by the inverse of the luminosity of the
trigger used to select it, such that the underlying dijet kinematic
distributions in the measurement correspond to the full two-jet
cross-section.

Events with two jets were selected, where the transverse momenta of
the jets were $p_\mathrm{T,1} > 50$~\GeV, $p_\mathrm{T,2} > 20$~\GeV,
and $\pTavg > 50$~\GeV. Both jets were required to have $\eta_{1,2} >
-2.8$ to separate them by $0.4$ units in pseudorapidity from the
\sumET-measuring region. Furthermore, the leading jet was also required to have
$\eta_{1} < 3.2$ to match the acceptance of the central-jet trigger.

For each selection on dijet kinematics, either as a function of \pTavg
and \etadijet or as a function of \tildexone and \tildextwo,
\meansumET was determined from the mean value of the \sumETcalib
distribution. The two values of \meansumET as measured in the forward
calorimeter at negative pseudorapidity and in the forward calorimeter
at positive pseudorapidity under the inverted-sign convention (see
Fig.~\ref{fig:Schematic}) were averaged to yield the presented
results.

The resolution on $p_\mathrm{T,1}$ and $p_\mathrm{T,2}$ and the
splitting of particle-level jets in the reconstruction resulted in a
migration of some events to adjacent \pTavg, \tildexone and \tildextwo
bins. This migration was corrected by applying a multiplicative factor
to the results. This factor was determined in simulation by taking the
ratio of the $\left<\smash{\sumETgen}\right>$ evaluated as a function
of reconstructed dijet variables to that evaluated with the jets at
the particle level. Since \PYTHIA 6 was found to best describe the jet
spectra and various jet-event-topology variables, it was used to
derive this bin-by-bin correction, which was typically only a few
percent from unity.

\section{Systematic uncertainties}

The results presented in this \paper\ are susceptible to several
sources of systematic uncertainty. The uncertainty from each source
was evaluated by analysing the data or deriving the corrections with a
corresponding variation in the procedure, averaging the \meansumET
results from each forward calorimeter side, and observing the changes
from the nominal results. The uncertainties from different sources
were treated as uncorrelated and added in quadrature to determine the
total uncertainty.

The \sumET calibration procedure is susceptible to uncertainties in
the overall energy scale of the forward calorimeter, in the amount of
material upstream of the calorimeter, in the physics model used to
derive it, and in the modelling of pileup in simulation. These
uncertainties were determined by deriving a new \sumET calibration for
each variation corresponding to a systematic uncertainty and applying
it to the data. To evaluate the energy scale uncertainty, the
calorimeter response in simulation was varied in an $\eta$-dependent
manner by an uncertainty derived from previous studies of
$\pi^0\to\gamma\gamma$ candidates in $\sqrt{s} = 7$~\TeV\ collision
data and simulation~\cite{Aad:2012mfa}, and from comparisons of
beam-test data with simulation~\cite{Pinfold:2012ac}. The resulting
changes in \meansumET from negative and positive variations of the
response were $+4$\% and $-8$\% respectively. To account for the
uncertainty in the amount of material upstream of the forward
calorimeter, the analysis described in Ref.~\cite{Aad:2012mfa}, which
evaluated the response in simulations with increased material in these
regions for $\sqrt{s} = 7$~\TeV\ events, was adapted to the conditions
of this analysis. Those results were used to vary the response in this
analysis, which resulted in changes of \meansumET by $\pm2$\%.

To evaluate the sensitivity to the physics model, \sumET calibrations
were derived using simulated \PYTHIA 6 and \HERWIG events and compared
to that derived using \PYTHIA 8. The variations among the three
generators in distributions relevant to the \sumET measurement, such
as the distribution of \sumET values,
$\mathrm{d}N/\mathrm{d}\!\sumET$, or the pseudorapidity distribution,
$\mathrm{d}\!\sumET/\mathrm{d}\eta$, were found to reasonably span
those in data. Thus, the largest difference in the results when using
the calibrations derived from any two generators, $5$\%, was
symmetrised and assigned as the uncertainty associated with the
sensitivity to the physics model.

The uncertainty in the modelling of the pileup within the simulation
was determined to be $\pm2$\% by investigating the sensitivity of the
\sumET calibration to several factors. These included varying the mean
number of $pp$ interactions per crossing, varying the pileup rejection
requirement, and accounting for possible mismodelling in the
simulation of the residual contribution to \sumET from unrejected
pileup vertices.

An additional uncertainty arising from possible defects in the
performance of the \sumET calibration was obtained from checking the
closure of the calibration procedure. The \sumET calibration, derived
from a \PYTHIA 8 event sample with a wide kinematic range, was found
to differ from unity when evaluated for subsets of the sample with
narrower selections on the dijet kinematics. In the simulation, this
behaviour results from a number of effects, such as the dependence of
the mean \ET per generator particle and the shape of the
$\mathrm{d}\!\sumET/\mathrm{d}\eta$ distribution on the selected dijet
kinematics, both of which affect the average response. A conservative
symmetric uncertainty of $5$\% was chosen to account for the potential
differences of the closure values from unity observed in simulation.

The uncertainty in the correction for bin migration effects, evaluated
by considering the sensitivity of the corrections to alternative
generators (\PYTHIA 8 and \HERWIG) and to variations in the jet energy
scale and resolution, was found to be smaller than
$\pm1$\%. Additional internal cross-checks on the \meansumET results
were investigated in the data. The nominal results were compared to an
alternative analysis in which the cells were combined into topological
clusters~\cite{Lampl:2008zz} and a new calibration was derived for the
detector-level \sumETraw constructed from the sum of cluster
transverse energies. An uncertainty of $\pm1$\% in the \meansumET was
assigned from this cross-check. Results determined using each side of
the forward calorimeter separately were compared and found to be
consistent. Additional potential sources of systematic uncertainty,
such as that in the energy resolution of the forward calorimeter, were
found to be negligible.

\begin{table}
\caption{\label{table:systematics} Relative systematic uncertainties
  for the measurements of \meansumET and \meansumET/\meansumETref,
  shown for each individual source of uncertainty. An entry of ``-''
  means that the source was found to be negligible.}
\begin{center}
{
\begin{tabular}{|l||c|c|}
\hline
& \multicolumn{2}{c|}{Typical uncertainty in} \\
\multicolumn{1}{|c||}{Source} & $\left<\sum\!{E_\mathrm{T}}\right>$ & $\left<\sum\!{E_\mathrm{T}}\right>$/$\left<\sum\!{E_\mathrm{T}}\right>^\mathrm{ref}$ \\ \hline \hline
$\sum\!{E_\mathrm{T}}$ calibration                          & & \\
\hspace{1.2cm} Forward calorimeter response & $+4$\%/$-8$\%  & - \\
\hspace{1.2cm} Extra material               & $\pm2$\%       & - \\
\hspace{1.2cm} Physics model                & $\pm5$\%       & $\pm1$\% \\
\hspace{1.2cm} Pileup modelling              & $\pm2$\%       & - \\
$\sum\!{E_\mathrm{T}}$ calibration closure                  & $\pm5$\%       & $\pm5$\% \\
Dijet kinematics bin migration              & $\pm1$\%       & $\pm1$\% \\
Calorimeter cluster cross-check             & $\pm1$\%       & $\pm1$\% \\ \hline
\multicolumn{1}{|l||}{Total uncertainty}    & $+9$\%/$-11$\% & $\pm5$\% \\
\hline
\end{tabular}
}
\end{center}
\end{table}

For most of the kinematic range except at high \pTavg, or when
$\tildexone$ or $\tildextwo$ is large, the statistical uncertainties
are negligible compared to the systematic ones. The dominant
uncertainties in the \meansumET measurement are from the energy scale,
the physics model, and the variation of the \sumET response with dijet
kinematics. The total uncertainty is $+9\%/$-$11\%$ and varies only
weakly with selections on dijet kinematics. The uncertainty in the
\meansumET/\meansumETref quantity was determined by varying the
numerator and denominator according to each source simultaneously to
properly account for their cancellation in the ratio. The resulting
uncertainty is $\pm5$\%, dominated by the variation of the \sumET
response with kinematics, which by its nature does not cancel in the
ratio of \meansumET for different kinematic selections. The total
systematic uncertainty is summarised in Table~\ref{table:systematics}
for the \meansumET and \meansumET/\meansumETref quantities.

A further cross-check was performed to determine the average
contribution to the mean \sumET from any additional jet in the
events. This contribution was estimated by repeating the analysis and
rejecting events with a $\pT > 15$~\GeV\ jet in $\eta < -2.8$, and was
found to be smaller than $2$\%. Since the \sumET definition includes
this energy, no uncertainty is assigned or correction applied.

\section{Results}

This section shows the \meansumET and \meansumET/\meansumETref
results, corrected to the particle level. In all distributions the
events are required to contain two particle-level jets with $p_\mathrm{T,1} >
50$~\GeV, $p_\mathrm{T,2} > 20$~\GeV, and $\pTavg > 50$~\GeV. Both
jets are required to have $\eta_{1,2} > -2.8$ and the leading jet is
also required to have $\eta_{1} < 3.2$.

\begin{figure}[!t]
\centering
\includegraphics[width=0.99\columnwidth]{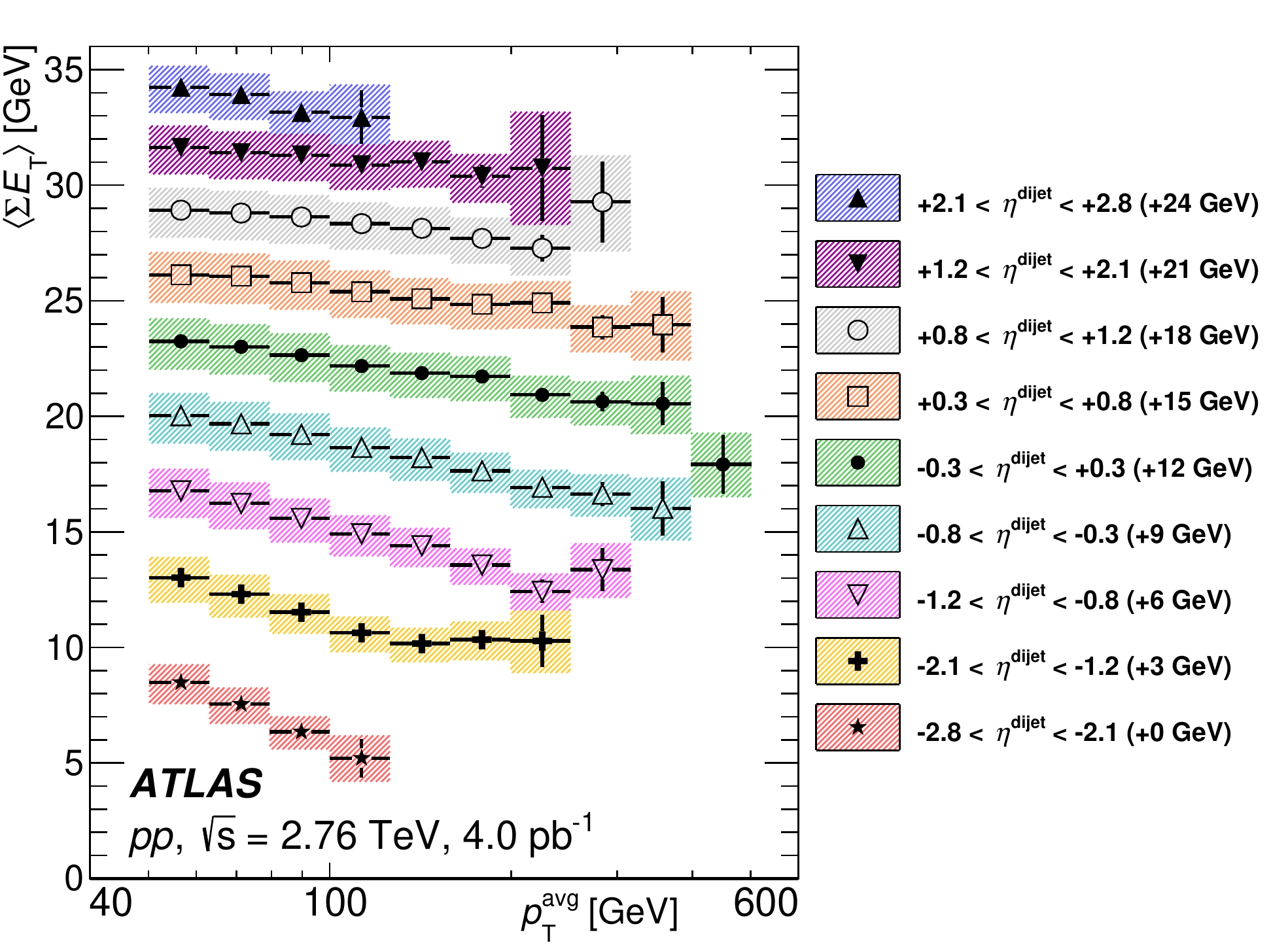}
\caption{\label{fig:aux0} Measured average sum of the transverse
  energy at large pseudorapidity \meansumET in hard-scatter
  \pp\ collisions, shown as a function of the average dijet momentum
  \pTavg. Each series depicts a different range of the average dijet
  pseudorapidity $\etadijet$ and the series are displaced vertically
  for clarity by the amount given in round brackets in the legend. The
  vertical shaded bands represent the total systematic and statistical
  uncertainties on the data in quadrature while the vertical bars
  represent statistical uncertainties only.}
\end{figure}

\begin{figure}[!t]
\centering
\includegraphics[width=0.6\columnwidth]{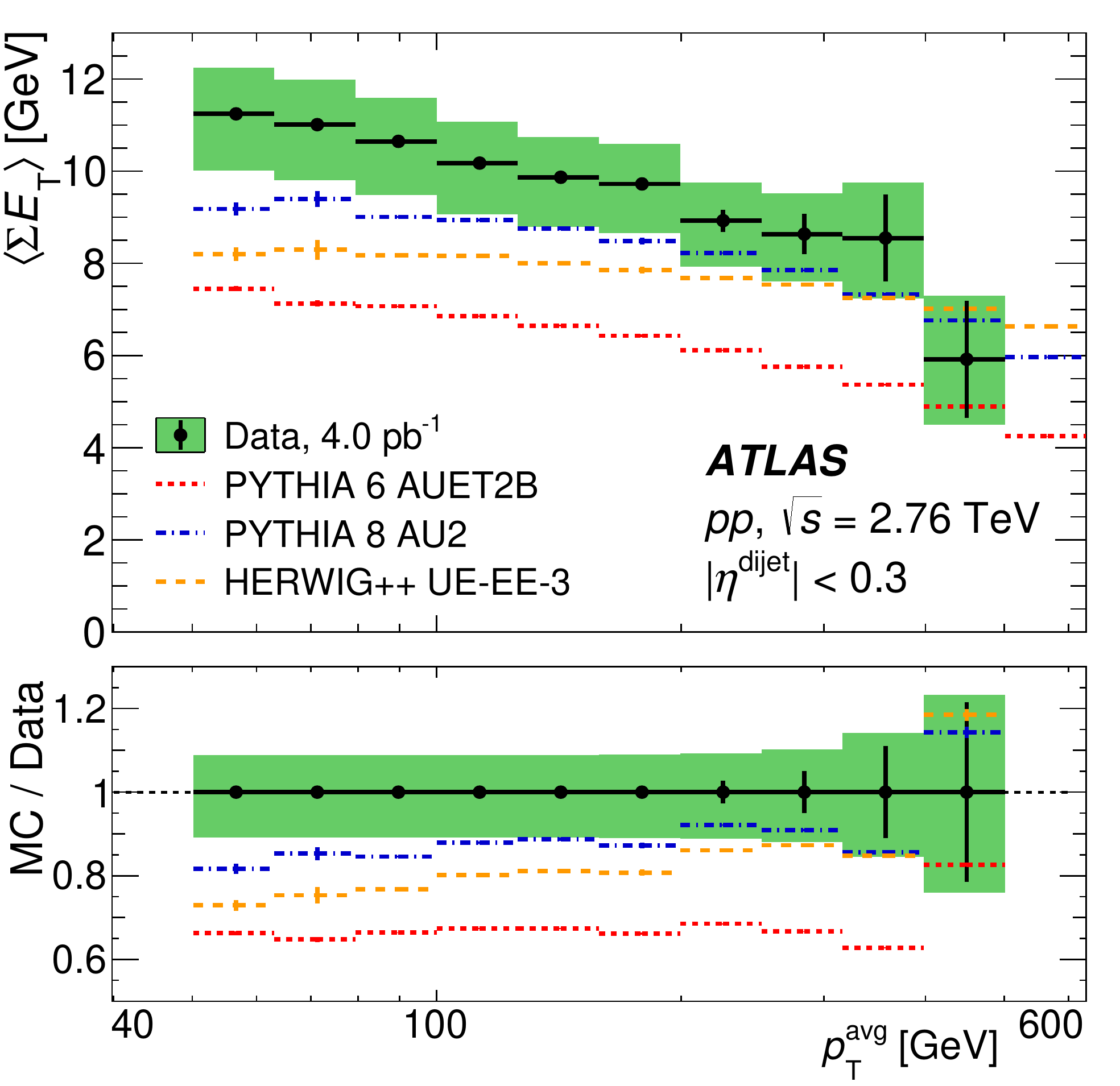}
\caption{\label{fig:OverallSumET} Measured \meansumET in hard-scatter
  \pp\ collisions, shown as a function of \pTavg for
  $\left|\smash{\etadijet}\right| < 0.3$ and in comparison with the
  predictions of three MC event generators. The vertical shaded bands
  represent total systematic and statistical uncertainties in the data
  in quadrature while the vertical bars represent statistical
  uncertainties only. The bottom panel shows the ratio of the
  predictions of the three MC generators to the data.}
\end{figure}

Fig.~\ref{fig:aux0} shows an overview of the measured \meansumET
values as a function of \pTavg for each range of \etadijet and
summarises the range of dijet kinematics accessed in the
measurement. Fig.~\ref{fig:OverallSumET} shows the \meansumET as a
function of \pTavg for central jet pairs
$\left(\left|\smash{\etadijet}\right| < 0.3\right)$ in more
detail. The \meansumET is anti-correlated with the dijet \pTavg,
decreasing by $25$\% as \pTavg varies from $50$~\GeV\ to
$500$~\GeV. The bottom panel of Fig.~\ref{fig:OverallSumET} shows the
ratio of the \meansumET in these generators to that in the
data. \PYTHIA 8 best reproduces \meansumET in data, typically agreeing
within one and a half times the uncertainty of the data in the
kinematic selections shown here and in most other selections
analysed. While the generators systematically underpredict the overall
scale of the \sumET production, the \meansumET is generally
anticorrelated with \pTavg in each one just as it is in the data. The
observation of an anticorrelation with \pTavg at mid-rapidity in $pp$
collisions is important for interpreting the \pPb results, since it
indicates a non-trivial correlation between hard-scattering kinematics
and \sumET production, but the \pTavg quantity offers only an indirect
relationship to the underlying Bjorken-$x$ values. The first point in
the upper panel of Fig.~\ref{fig:OverallSumET} shows the reference
value for data of $\meansumETref = 11.2^{+1.0}_{-1.2}$~\GeV. For the
generators considered in this analysis, the value of \meansumETref in
simulation is $7.5$~\GeV\ in \PYTHIA 6, $9.2$~\GeV\ in \PYTHIA 8, and
$8.2$~\GeV\ in \HERWIG.

\begin{figure}[!t]
\centering
\includegraphics[width=0.6\columnwidth]{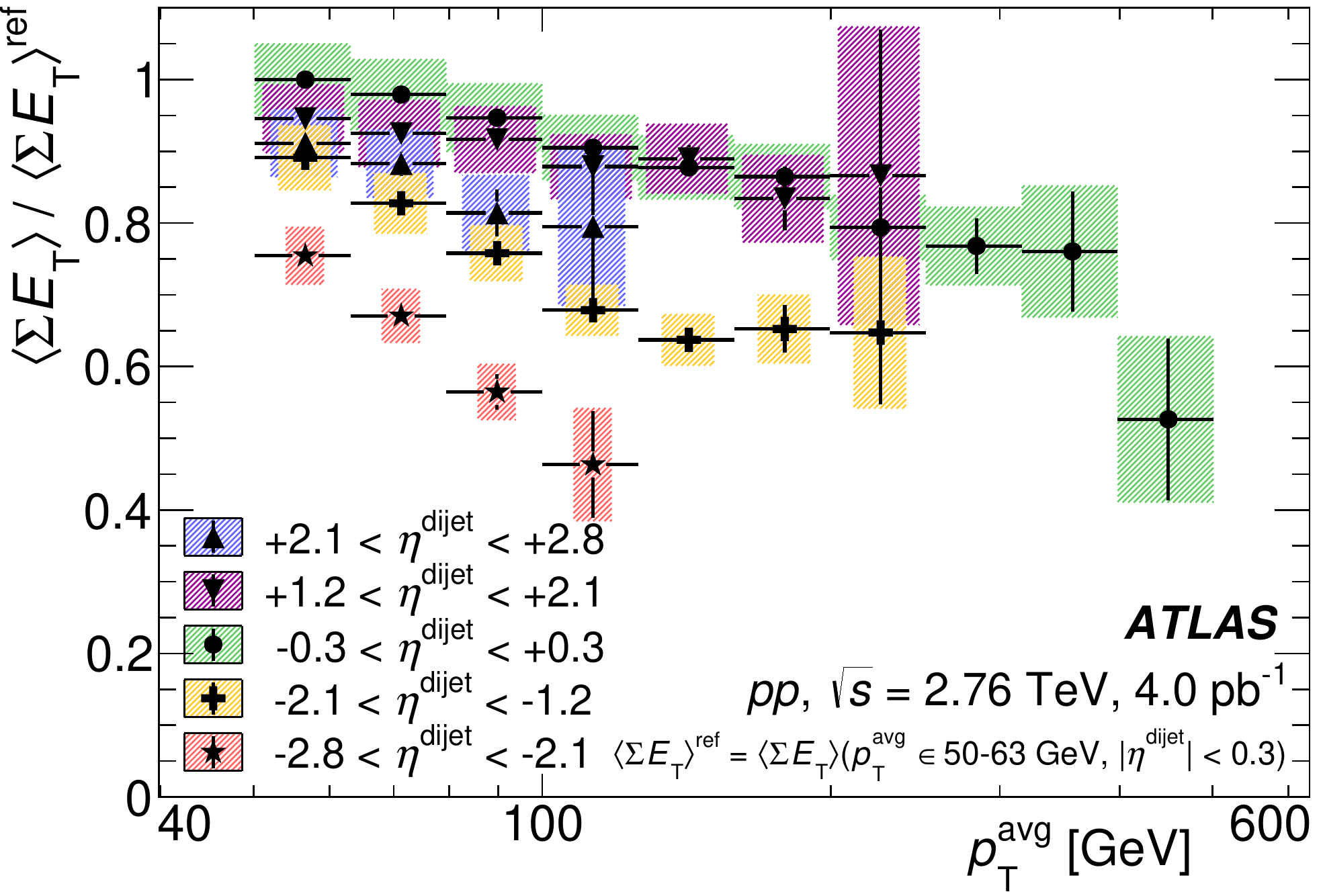}
\caption{\label{fig:RelativeSumET_pTavg_etadijet} Measured ratio
  $\meansumET/\meansumETref$ in hard-scatter \pp collisions, shown as
  a function of \pTavg for different selections on \etadijet. The
  vertical shaded bands represent the total systematic and statistical
  uncertainties in quadrature while the vertical error bars represent
  statistical uncertainties only. When two error bands overlap
  vertically, their horizontal widths have been adjusted so that the
  edges of both are visible. }
\end{figure}

To further explore the variation of the results with the Bjorken-$x$
of the hard-scattered partons, the dependence on the average
pseudorapidity of the dijet was investigated. At fixed \pTavg, dijets
with large positive or negative \etadijet arise from parton--parton
configurations with large $x$ in the projectile or target proton,
respectively. Fig.~\ref{fig:RelativeSumET_pTavg_etadijet} shows the
ratio \meansumET/\meansumETref as a function of \pTavg for different
ranges of \etadijet. In the ratio \meansumET/\meansumETref, much of
the uncertainty in the data and the overall scale difference between
data and the generators cancels, allowing a precise measurement of the
relative dependence of \meansumET on dijet kinematics and comparison
to generators. When the dijet pair is at positive pseudorapidity (in
the direction of the projectile proton), the relationship between the
\meansumET and \pTavg is similar to that for mid-rapidity
dijets. However, as the dijet pair pseudorapidity moves to negative
rapidities close to the \sumET-measuring region (in the direction of
the target proton), this anti-correlation becomes stronger and the
overall level of the \sumET decreases. For the \etadijet selection
nearest to the region in which the \sumET is measured ($-2.8 <
\etadijet < -2.1$), \meansumET decreases by $40$\% as \pTavg increases
by a factor of two from $50$~\GeV\ to $100$~\GeV.

\begin{figure}[!t]
\centering
\includegraphics[width=0.98\columnwidth]{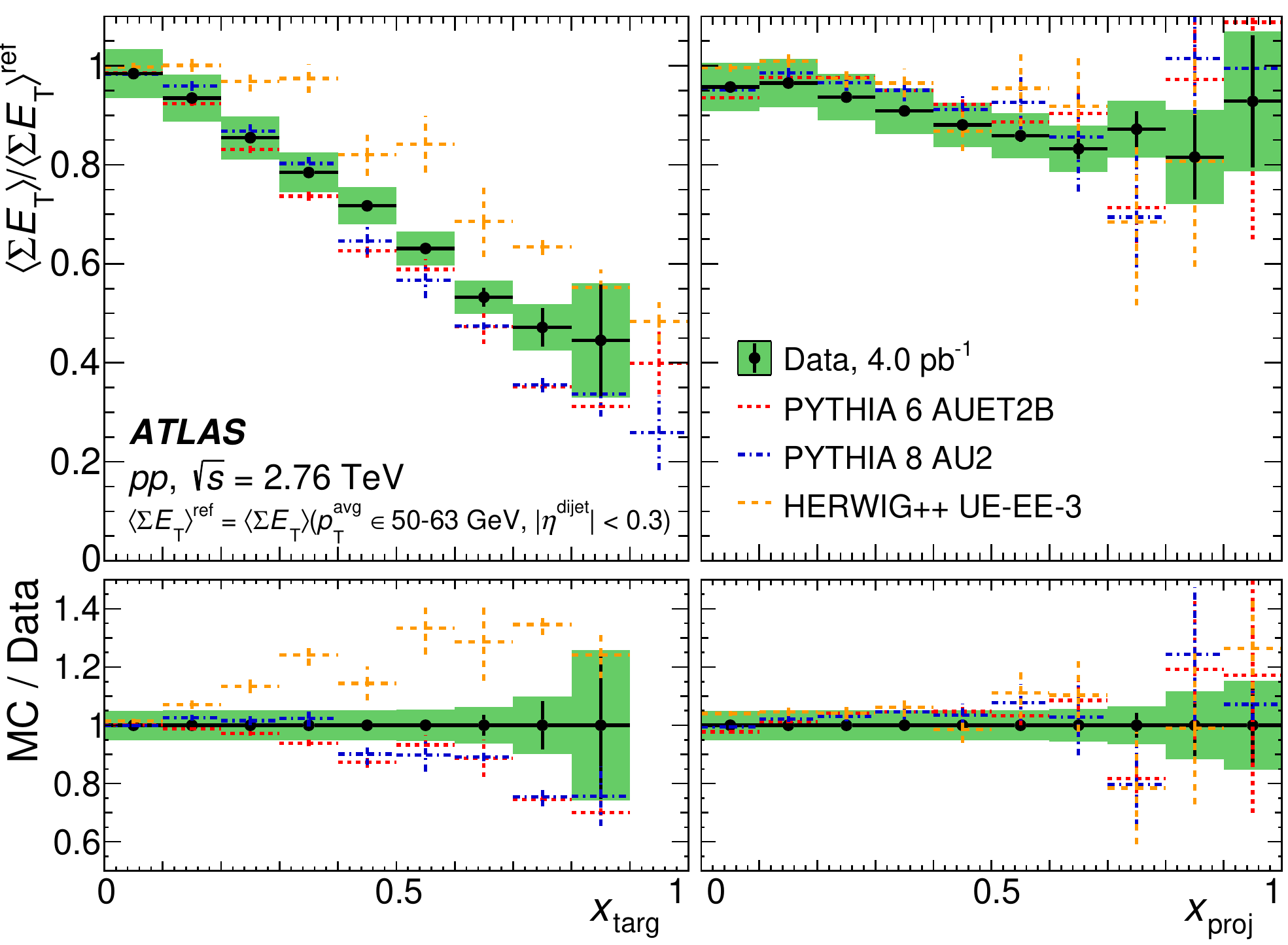}
\caption{\label{fig:RelativeSumET_x1_x2} Measured ratio
  $\meansumET/\meansumETref$ in hard-scatter \pp\ collisions, shown as
  a function of \tildextwo (left) and \tildexone (right). The vertical
  shaded bands represent total systematic and statistical
  uncertainties in the data in quadrature while the vertical bars
  represent statistical uncertainties only. The bottom panel shows the
  ratio of the predictions of three MC event generators to the data.}
\end{figure}

Finally, the pattern of how the \meansumET values for dijets at all
\pTavg and \etadijet depend on the underlying hard scattering
kinematics can be explored more directly by plotting them as a
function of the kinematic variables \tildexone and
\tildextwo. Fig.~\ref{fig:RelativeSumET_x1_x2} shows the ratio
\meansumET/\meansumETref as a function of each variable, while
integrating over the other. The value of \meansumET is largely
insensitive to \tildexone (which corresponds to the Bjorken-$x$ in the
proton moving to positive rapidity), changing by only $10$\% over the
entire range $0 < \tildexone < 1$. On the other hand, \meansumET
varies strongly with \tildextwo (which corresponds to the Bjorken-$x$
in the proton moving to negative rapidity), decreasing by more than a
factor of two between $\tildextwo = 0$ and $0.9$ in an approximately
linear fashion.

Since \tildexone and \tildextwo are generally anti-correlated in dijet
events, the data were also analysed by fixing each variable in a
narrow range and testing the dependence of \meansumET on the other,
and this gave results quantitatively similar to those in
Fig.~\ref{fig:RelativeSumET_x1_x2}. The generators considered here
have qualitatively similar behaviour. They describe the \tildexone
dependence well, but \PYTHIA 6 and \PYTHIA 8 show a slightly stronger
dependence on \tildextwo, while \HERWIG shows a much weaker one. The
observed dependence admits a simple interpretation: when the hard
scattering involves a parton with large \tildextwo, the beam remnant
has less longitudinal energy and transverse energy production at large
pseudorapidity is substantially reduced.

\FloatBarrier

\section{Conclusions}

This \paper\ presents measurements of the dependence of transverse
energy production at large rapidity on hard-scattering kinematics in
$4.0$~pb$^{-1}$ of $\sqrt{s}=2.76$~\TeV\ \pp\ collision data with the
ATLAS detector at the LHC. The results have a number of
implications. They demonstrate that the average level of transverse
energy production at large pseudorapidity is sensitive mainly to the
Bjorken-$x$ of the parton originating in the beam-proton which is
headed towards the energy-measuring region, and is largely insensitive
to $x$ in the other proton. Specifically, the decrease in the mean
transverse energy downstream of a beam-proton is approximately linear
in the longitudinal energy carried away from that beam-proton in the
hard scattering. Monte Carlo event generators generally underpredict
the overall value of the transverse energy but properly model with
varying accuracy the trend in how this quantity depends on
hard-scattering kinematics.

These results provide counter-evidence to claims that the observed
centrality-dependence of the jet rate in \pPb\ collisions simply
arises from the suppression of transverse energy production at
negative rapidity in the hard-scattered $NN$ sub-collision. In the
\pPb data, the deviations from the expected centrality dependence are
observed to depend only on, and increase with, $x$ in the
proton. Therefore, for this effect to be consistent with arising from
a feature of $NN$ collisions, transverse energy production at small
angles should decrease strongly and continuously with increasing $x$
in the proton headed in the opposite direction (corresponding to
\tildexone in this measurement). The results presented in this
\paper\ do not obviously support such a scenario.

In conclusion, the measurements presented in this \paper\ seek to
reveal the correlation between hard-process kinematics and transverse
energy production at large pseudorapidity which is present in
individual nucleon--nucleon collisions. As a \pPb collision can be
understood as a superposition of such interactions, the measurements
presented here may serve as a limiting case against which to test
descriptions of the underlying physics of hard and soft particle
production in \pPb collisions.

\section*{Acknowledgements}



We thank CERN for the very successful operation of the LHC, as well as
the support staff from our institutions without whom ATLAS could not
be operated efficiently.

We acknowledge the support of ANPCyT, Argentina; YerPhI, Armenia; ARC,
Australia; BMWFW and FWF, Austria; ANAS, Azerbaijan; SSTC, Belarus;
CNPq and FAPESP, Brazil; NSERC, NRC and CFI, Canada; CERN; CONICYT,
Chile; CAS, MOST and NSFC, China; COLCIENCIAS, Colombia; MSMT CR, MPO
CR and VSC CR, Czech Republic; DNRF, DNSRC and Lundbeck Foundation,
Denmark; IN2P3-CNRS, CEA-DSM/IRFU, France; GNSF, Georgia; BMBF, HGF,
and MPG, Germany; GSRT, Greece; RGC, Hong Kong SAR, China; ISF, I-CORE
and Benoziyo Center, Israel; INFN, Italy; MEXT and JSPS, Japan; CNRST,
Morocco; FOM and NWO, Netherlands; RCN, Norway; MNiSW and NCN, Poland;
FCT, Portugal; MNE/IFA, Romania; MES of Russia and NRC KI, Russian
Federation; JINR; MESTD, Serbia; MSSR, Slovakia; ARRS and MIZ\v{S},
Slovenia; DST/NRF, South Africa; MINECO, Spain; SRC and Wallenberg
Foundation, Sweden; SERI, SNSF and Cantons of Bern and Geneva,
Switzerland; MOST, Taiwan; TAEK, Turkey; STFC, United Kingdom; DOE and
NSF, United States of America. In addition, individual groups and
members have received support from BCKDF, the Canada Council, CANARIE,
CRC, Compute Canada, FQRNT, and the Ontario Innovation Trust, Canada;
EPLANET, ERC, FP7, Horizon 2020 and Marie Sk\l{}odowska-Curie Actions,
European Union; Investissements d'Avenir Labex and Idex, ANR, Region
Auvergne and Fondation Partager le Savoir, France; DFG and AvH
Foundation, Germany; Herakleitos, Thales and Aristeia programmes
co-financed by EU-ESF and the Greek NSRF; BSF, GIF and Minerva,
Israel; BRF, Norway; the Royal Society and Leverhulme Trust, United
Kingdom.

The crucial computing support from all WLCG partners is acknowledged
gratefully, in particular from CERN and the ATLAS Tier-1 facilities at
TRIUMF (Canada), NDGF (Denmark, Norway, Sweden), CC-IN2P3 (France),
KIT/GridKA (Germany), INFN-CNAF (Italy), NL-T1 (Netherlands), PIC
(Spain), ASGC (Taiwan), RAL (UK) and BNL (USA) and in the Tier-2
facilities worldwide.


\bibliographystyle{atlasBibStyleWoTitle}
\bibliography{ppDijetFwdSumETPaper}

\clearpage 
\input{atlas_authlist}

\end{document}

%% file: ppDijetFwdSumETPaper-metadata.tex
\AtlasTitle{Measurement of the dependence of
transverse energy production at large pseudorapidity on the
hard-scattering kinematics of proton--proton collisions at $\sqrt{s} =
2.76$~\TeV\ with ATLAS}

\usepackage{authblk}

\PreprintIdNumber{CERN-PH-EP-2015-293}

\AtlasJournal{Phys.\ Lett.\ B}

\AtlasAbstract{
The relationship between jet production in the central region and the
underlying-event activity in a pseudorapidity-separated region is
studied in $4.0$~pb$^{-1}$ of $\sqrt{s} = 2.76$~\TeV\ \pp\ collision
data recorded with the ATLAS detector at the LHC. The underlying event
is characterised through measurements of the average value of the sum
of the transverse energy at large pseudorapidity downstream of one of
the protons, which are reported here as a function of hard-scattering
kinematic variables. The hard scattering is characterised by the
average transverse momentum and pseudorapidity of the two highest
transverse momentum jets in the event. The dijet kinematics are used
to estimate, on an event-by-event basis, the scaled longitudinal
momenta of the hard-scattered partons in the target and projectile
beam-protons moving toward and away from the region measuring
transverse energy, respectively. Transverse energy production at large
pseudorapidity is observed to decrease with a linear dependence on the
longitudinal momentum fraction in the target proton and to depend only
weakly on that in the projectile proton. The results are compared to
the predictions of various Monte Carlo event generators, which
qualitatively reproduce the trends observed in data but generally
underpredict the overall level of transverse energy at forward
pseudorapidity.}

%% file: atlas_authlist.tex
\begin{flushleft}
{\Large The ATLAS Collaboration}

\bigskip

G.~Aad$^\textrm{\scriptsize 85}$,
B.~Abbott$^\textrm{\scriptsize 113}$,
J.~Abdallah$^\textrm{\scriptsize 151}$,
O.~Abdinov$^\textrm{\scriptsize 11}$,
R.~Aben$^\textrm{\scriptsize 107}$,
M.~Abolins$^\textrm{\scriptsize 90}$,
O.S.~AbouZeid$^\textrm{\scriptsize 158}$,
H.~Abramowicz$^\textrm{\scriptsize 153}$,
H.~Abreu$^\textrm{\scriptsize 152}$,
R.~Abreu$^\textrm{\scriptsize 116}$
% ...
\bigskip
\\
% ...
$^{11}$ Institute of Physics, Azerbaijan Academy of Sciences, Baku, Azerbaijan\\
% ...
$^{85}$ CPPM, Aix-Marseille Universit{\'e} and CNRS/IN2P3, Marseille, France\\
% ...
$^{90}$ Department of Physics and Astronomy, Michigan State University, East Lansing MI, United States of America\\
% ...
$^{107}$ Nikhef National Institute for Subatomic Physics and University of Amsterdam, Amsterdam, Netherlands\\
% ...
$^{113}$ Homer L. Dodge Department of Physics and Astronomy, University of Oklahoma, Norman OK, United States of America\\
% ...
$^{116}$ Center for High Energy Physics, University of Oregon, Eugene OR, United States of America\\
% ...
$^{151}$ Institute of Physics, Academia Sinica, Taipei, Taiwan\\
$^{152}$ Department of Physics, Technion: Israel Institute of Technology, Haifa, Israel\\
$^{153}$ Raymond and Beverly Sackler School of Physics and Astronomy, Tel Aviv University, Tel Aviv, Israel\\
% ...
$^{158}$ Department of Physics, University of Toronto, Toronto ON, Canada\\
% ...
\end{flushleft}
